\documentclass[10pt,twocolumn,english,conference]{IEEEtran}
\usepackage{multicol}
\usepackage{amsmath}
\usepackage{subeqnarray}
\usepackage{cases}
\usepackage{framed}
\usepackage{amsfonts}
\usepackage{amssymb}
\usepackage{graphicx}
\usepackage{subfigure}
\usepackage{babel}
\usepackage{cite}
\usepackage{color}%

\providecommand{\U}[1]{\protect\rule{.1in}{.1in}}

\newtheorem{proposition}{Proposition}

\begin{document}

\title{Sum-Rate Optimization in a Two-Way Relay Network with Buffering}
\author{%

\begin{tabular}
[c]{c}%
Huaping Liu$^{\dagger,\ast}$, Petar~Popovski$^{\ast}$, Elisabeth de Carvalho$^{\ast}$ and Yuping Zhao$^{\dagger}$\\
$\dagger$State Key Laboratory of Advanced Optical Communication Systems and Networks, Peking University, China\\
$\ast$Department of Electronic Systems, Aalborg University, Denmark\\
Email:  $\left\{ {{\text{liuhp,yuping.zhao}}} \right\}$@pku.edu.cn, $\left\{ {{\text{petarp,edc}}} \right\}$@es.aau.dk\\
\end{tabular}
}
\maketitle

\begin{abstract}
A Relay Station (RS) uses a buffer to store and process the received data packets before forwarding them. Recently, the buffer has been exploited in one-way relaying to opportunistically schedule the two different links according to their channel quality. The intuition is that, if the channel to the destination is poor, then
RS stores more data from the source, in order to use it when the channel to the destination is good. We apply this intuition to the case of half-duplex two-way relaying, where the interactions among the buffers and the links become more complex. We  investigate the sum-rate maximization problem in the Time Division Broadcast (TDBC): the users send signals to the RS in different time slots, the RS decodes and stores messages in the buffers. For downlink transmission, the RS re-encodes and sends using the optimal broadcast strategy.
The operation in each time slot is not  determined in advance, but depends on the channel state information (CSI). We derive the decision function for adaptive link selection with respect to CSI using the Karush-Kuhn-Tucker (KKT) conditions. The thresholds of the decision function are obtained under Rayleigh fading channel conditions. The numerical results show that the sum-rate of the adaptive link selection protocol with buffering is significantly larger compared to the reference protocol with fixed transmission schedule.
\end{abstract}

\section{Introduction}
Wireless networks with relays are a subject of intense research interest. Capacity bounds and various cooperative strategies for relay networks have been studied in \cite{Cover}. \cite{Kramer} developed Decode-and-Forward (DF) relaying to multiple access relay channels and broadcast relay channels, and generalized Compress-and-Forward (CF) relaying to multiple relays. A paradigm shift in communicating multiple flows through relays occurred with the concept of network coding \cite{Ahlswede}, where a relay transmits functions of the incoming communication flows, rather than only replicating the incoming flows. This idea has a particularly promising application in wireless networks, where it was shown that two-way relaying can be improved when the relay uses wireless network coding \cite{Popovski}, \cite{Katti}.

In an information-theoretic framework, scheduling of the different links of a relay system is usually not questioned. For example, in a one way relaying, the source first sends its data to the relay station (RS) and the RS forwards it. This operation is repeated sequentially. In a practical system, the relay needs a buffer to store the received packets, in order to process them before forwarding. Such a buffering gives an opportunity for a clever scheduling that exploits the channel state information (CSI), as proposed in \cite{ZlatanovPETAR_GLOBECOM}. If the link from the RS to the destination is weak, it might be beneficial not to forward any data but instead accumulate more data from the source, and wait for a better channel to forward any data. The advantage of a scheduling at the relay that is not fixed in advance was pointed out in \cite{Kramer04}, where it is proven that a protocol with random schedule is better than a protocol with a fixed schedule. The works \cite{Nikola} and \cite{Zlatanov} treat the one-way relaying scenario and reveal that buffer-aided protocols determined by instantaneous CSI outperforms conventional relaying protocols with fixed transmission time slots.

\begin{figure}[t]
\centering
\includegraphics[width=8 cm]{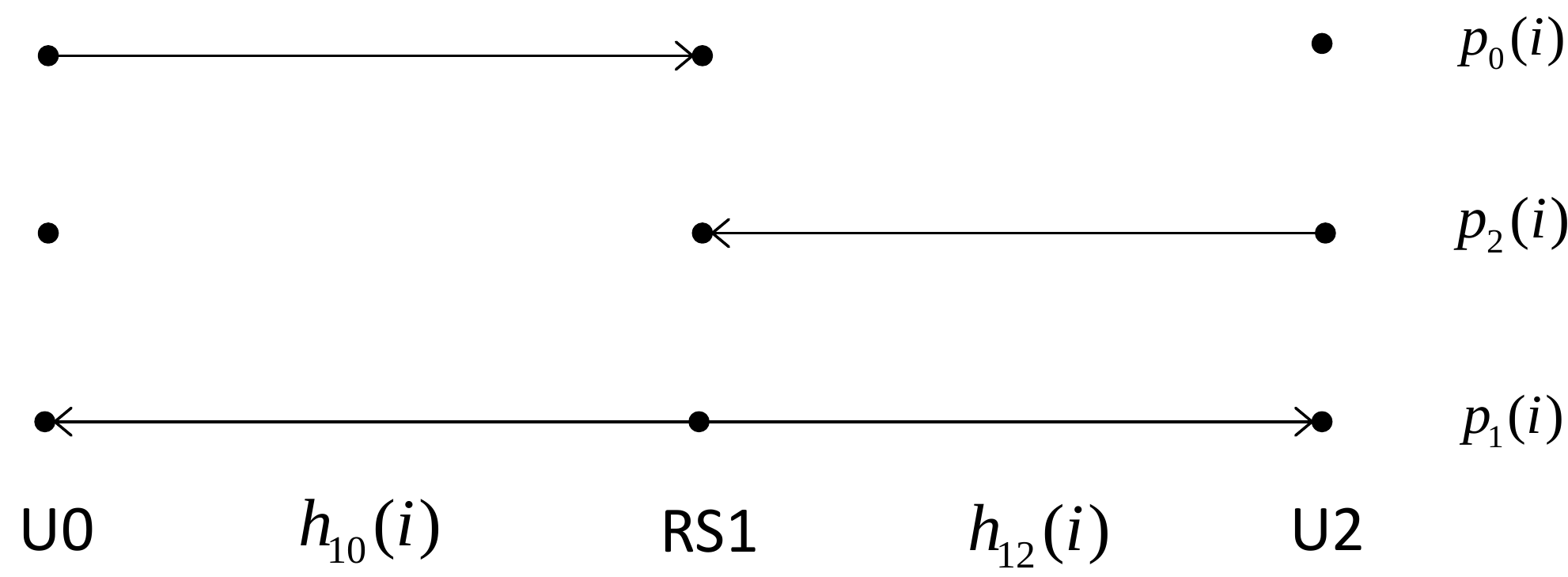}
\caption{TDBC two-way transmission model}
\label{system}%
\vspace{-5pt}
\end{figure}

In this paper we generalize the concept of buffer-aided relaying to the case of two-way relaying. The presence of two communication flows and buffers significantly changes the optimization problem. We focus on the sum-rate maximization problem in the Time Division Broadcast (TDBC) two-way relay network with buffering as shown in Fig.~\ref{system}. We use the optimal broadcast strategy introduced in \cite{Oechtering} for two-way relaying with side information at the terminals. We solve the optimization problem by relaxing the discrete constraints and applying the Karush-Kuhn-Tucker (KKT) condition. The numerical result shows that the sum-rate of the proposed adaptive link selection protocol with buffering significantly exceeds the reference protocol with fixed transmission schedule.

\section{System Model}\label{model}
Consider a two-way relay network in which two users (U$0$ and U$2$) intend to exchange information with the aid of a relay station (RS$1$) as Fig.~\ref{system} shows. RS$1$ applies Decode-and-Forward (DF) and it has two buffers for storing messages, one for each of the users, assumed to be of unlimited size.
All the nodes are half-duplex, such that a node can either transmit or receive at a given time. We assume that there is no direct link between U$0$ and U$2$. The elementary transmission unit is a time slot of fixed duration. We assume a block fading channel, such that the channel is constant over the duration of one time slot but changes independently from one slot to another.
 ${h_{01}}(i)$ and ${h_{12}}(i)$ are the channels of link U$0$$\rightarrow$RS$1$ and RS$1$$\rightarrow$U$2$ in time slot $i$ respectively. In every time slot $i$, all the nodes know the instantaneous channels which are assumed to be reciprocal. $P$ is the transmission power for each node. The noise at all receivers ${z_j} \sim \mathcal{CN}(0,{\sigma ^2}) , j \in \{ 0,1,2\}$ is an independent Additive White Gaussian Noise (AWGN) with zero mean and variance $\sigma^2$. The instantaneous Signal-to-Noise Ratio (SNR) for link U$0$$ \leftrightarrow $RS$1$ and RS$1$$ \leftrightarrow $U$2$ is ${\gamma _0}\left( i \right) = {{P{{\left| {{h_{01}}\left( i \right)} \right|}^2}} \over {{\sigma ^2}}}$ and ${\gamma _2}\left( i \right) = {{P{{\left| {{h_{12}}\left( i \right)} \right|}^2}} \over {{\sigma ^2}}}$ respectively. The expected value of the SNR is ${\Omega _l} = E\left\{ {{\gamma _l}(i)} \right\}, l \in \left\{ {0,2} \right\}$. When user $l$ transmits in slot $i$, the maximal instantaneous rate is ${C_l}\left( i \right) = {\log _2}\left( {1 + {\gamma _l}\left( i \right)} \right)$. The transmission from RS$1$ is a broadcast process with side information at two terminals (U$0$ and U$2$). From \cite{Oechtering}, the maximal broadcast rate for each link is ${C_0}\left( i \right)$ and ${C_2}\left( i \right)$, i.e. the maximal instantaneous achievable rate for each individual link.

\section{Transmission protocol design}
\subsection{Operation of the buffers}
RS$1$ has two buffers $Q_0$ and $Q_2$ to store the decoded messages from U$0$ and U$2$ respectively.  At a specific time slot $i$, only one node 0, 1 and 2 has the possibility to transmit a signal. The transmission schedule is determined at each time slot and depends on the instantaneous CSI and the buffers' state. The variable ${p_j}\left( i \right) \in \left\{ {0,1} \right\}$ indicates whether node $j \in \left\{ {0,1,2} \right\}$ transmits in time slot $i$ (${p_j}\left( i \right) = 1$) or not (${p_j}\left( i \right) = 0$). Obviously, ${p_0}(i)  + {p_1}(i)  + {p_2}(i) = 1$ has to be satisfied.
In slot $i$, the buffers are updated as follows.
\begin{itemize}
\item If U$0$ transmits:
\begin{equation}
\vspace{-3pt}
{Q_0}\left( i \right) = {Q_0}\left( {i - 1} \right) + {C_0}\left( i \right),~  {Q_2}\left( i \right) = {Q_2}\left( {i - 1} \right).
\label{bufferu1}
\end{equation}

\item If U$2$ transmits:
\begin{equation}
\vspace{-3pt}
{Q_0}\left( i \right) = {Q_0}\left( {i - 1} \right),~  {Q_2}\left( i \right) = {Q_2}\left( {i - 1} \right) + {C_2}\left( i \right).
\label{bufferu2}
\end{equation}

\item If RS$1$ transmits:
\begin{subequations}
\vspace{-3pt}
\label{bufferrs1}
\begin{align}
\label{Q0rs1} &{Q_0}\left( i \right) = {Q_0}\left( {i - 1} \right) - \min \left\{ {{C_2}\left( i \right),{Q_0}\left( {i - 1} \right)} \right\}\\
\label{Q2rs1} &{Q_2}\left( i \right) = {Q_2}\left( {i - 1} \right) - \min \left\{ {{C_0}\left( i \right),{Q_2}\left( {i - 1} \right)} \right\}.
\end{align}
\end{subequations}
\end{itemize}

The optimal transmission strategy should satisfy:
\begin{equation}
E\left\{ {{Q_l}\left( i \right) - {Q_l}\left( {i - 1} \right)} \right\} = 0,~ l=1,2
\label{bufferucon1}
\end{equation}
indicating that the number of bits in each buffer should be stable for optimal operation.

Incorporating (\ref{bufferu1})-(\ref{bufferucon1}) and considering the transmission indicator ${p_j}(i)$ of each node, we get the criterion for optimal transmission as follow
\begin{subequations}
\label{p0p1p2}
\begin{align}
\label{p0p1} &E\left\{ {{p_0}(i){C_0}\left( i \right)} \right\} = E\left\{ {{p_1}(i)\min \left\{ {{C_2}\left( i \right),{Q_0}\left( {i - 1} \right)} \right\}} \right\}\\
\label{p2p2} &E\left\{ {{p_2}(i){C_2}\left( i \right)} \right\} = E\left\{ {{p_1}(i)\min \left\{ {{C_0}\left( i \right),{Q_2}\left( {i - 1} \right)} \right\}} \right\}
\end{align}
\end{subequations}

Based on (\ref{p0p1p2}) and similar to \cite[Theorem 2]{Nikola}, the impact of ${C_2}\left( i \right) >  {Q_0}\left( {i - 1} \right)$ and ${C_0}\left( i \right) >  {Q_2}\left( {i - 1} \right)$ is negligible over a very long period and (\ref{p0p1p2}) becomes
\begin{subequations}
\vspace{-3pt}
\label{p0p1p2s2}
\begin{align}
\label{p0p1s2} &E\left\{ {{p_0}(i){C_0}\left( i \right)} \right\} = E\left\{ {{p_1}(i){C_2}\left( i \right)} \right\}\\
\label{p2p2s2} &E\left\{ {{p_2}(i){C_2}\left( i \right)} \right\} = E\left\{ {{p_1}(i){C_0}\left( i \right)} \right\}.
\end{align}
\vspace{-15pt}
\end{subequations}

\subsection{KKT condition for maximizing the system sum-rate}
Our goal is to maximize the average sum-rate of the two-way relaying system
under the buffer stability conditions (\ref{p0p1p2s2}).
We consider an observation window of $N$ time slots for which we define the target function $f\left( \mathbf{p} \right)$ and the constraint functions $h_k\left( \mathbf{p} \right), k \in 0,~ k = 1,...,N + 2$  in (\ref{fh}). $N$ is assumed to be asymptotically large so that, assuming ergodicity, the time average is equivalent to the ensemble average:
\begin{subequations}
\label{fh}
\begin{align}
\label{f} &f\left( \mathbf{p} \right) = {1 \over N}\sum\limits_{i = 1}^N {{p_1}(i)\left[ {{C_2}\left( i \right) + {C_0}\left( i \right)} \right]}\\
\label{h1} &{h_1}\left( \mathbf{p} \right) = {1 \over N}\sum\limits_{i = 1}^N {\left[ {{p_0}(i){C_0}\left( i \right) - {p_1}(i){C_2}\left( i \right)} \right]}\\
\label{h2} &{h_2}\left( \mathbf{p} \right) = {1 \over N}\sum\limits_{i = 1}^N {\left[ {{p_2}(i){C_2}\left( i \right) - {p_1}(i){C_0}\left( i \right)} \right]}\\
\label{h3n} &{h_{2 + i}}\left( \mathbf{p} \right) = 1 - {p_0}(i) - {p_1}(i) - {p_2}(i)
\end{align}
\end{subequations}
here $\mathbf{p} = \left( {{p_0}(1),{p_1}(1),{p_2}(1),...,{p_0}(N),{p_1}(N),{p_2}(N)} \right)$.

Our optimization problem consists in
maximizing the average sum-rate over the transmission indicators ${p_j}(i)$:
\begin{eqnarray}
\nonumber
\vspace{-3pt}
\label{max} &&\mathop {\max }\limits_{{p_j}(i)}  ~~f\left( \mathbf{p} \right)  \hfill \\
\label{conh} s.t.&&{h_k}\left( \mathbf{p} \right) = 0,~ k = 1,...,N + 2,~~ N \to \infty \hfill \\\nonumber
\label{conp} &&{p_j}(i)\left[ {1 - {p_j}(i)} \right] = 0,~j \in \left\{ {0,1,2} \right\},~ i \in \left\{ {1,...,N} \right\} \nonumber
\vspace{-3pt}
\end{eqnarray}

Note that ${p_j}(i)$ is binary, so the  optimization problem above is over a discrete domain. We therefore relax the constraints, by assuming that ${p_j}(i)$ takes continuous values within $[0,1]$. In this new formulation, the constraint ${p_j}(i)\left[ {1 - {p_j}(i)} \right] = 0$ is replaced by $0 \le {p_j}(i) \le 1$. We can now resort to the KKT conditions.
Equivalence of this formulation is proved subsequently where
we show that the optimal points are on the border of the region $[0,1]$ which  coincides with the fact  that
the ${p_j}(i)$ takes binary values. The optimization problem is formulated as follows:
\begin{subequations}
\vspace{-10pt}
\label{kktcon}
\begin{align}
\label{kkt1} \nonumber &\nabla f\left( {{\mathbf{p}^ * }} \right) - \lambda \nabla {h_1}\left( {{\mathbf{p}^ * }} \right) - \mu \nabla {h_2}\left( {{\mathbf{p}^ * }} \right) - \sum\limits_{i = 1}^N {{\alpha _i}\nabla {h_{2 + i}}\left( {{\mathbf{p}^ * }} \right)} + \\
& \sum\limits_{j = 0}^2 {\sum\limits_{i = 1}^N {\beta _i^{\left( j \right)}\nabla \left[ {1 - p_j^*(i)} \right]} } + \sum\limits_{j = 0}^2 {\sum\limits_{i = 1}^N {\psi _i^{\left( j \right)}\nabla p_j^*(i)} }  = 0   \\
\label{kkt2} &0 \le p_j^*(i) \le 1,~j \in \left\{ {0,1,2} \right\} \\
\label{kkt3} &\beta _i^{\left( j \right)}\left[ {1 - p_j^*(i)} \right] = 0 ,~   \beta _i^{\left( j \right)} \ge 0  \\
\label{kkt4} &\psi _i^{\left( j \right)}p_j^*(i) = 0 ,~   \psi _i^{\left( j \right)} \ge 0 \\
\label{kkt5} &{h_{2+i}}\left( {{\mathbf{p}^ * }} \right) = 0,~ i = 1,...,N,~ N \to \infty \\
\label{kkt6} &{h_k}\left( {{\mathbf{p}^ * }} \right) = 0,~ k = 1,2
\end{align}
\end{subequations}

The KKT necessary conditions state that if ${{\mathbf{p}^ * }}$ is a local optimum, there exist constant coefficients $\lambda$, $\mu $, ${{\alpha _i}}$, ${\beta _i^{\left( j \right)}}$ and ${\psi _i^{\left( j \right)}}$ such that (\ref{kktcon}) is satisfied. Furthermore, because the target function and constraint functions in (\ref{kktcon}) are linear, the KKT necessary conditions are also sufficient conditions and a local maximal point is  the global maximal point as well. This means, if there exists  constant coefficients $\lambda$, $\mu $, ${{\alpha _i}}$, ${\beta _i^{\left( j \right)}}$ and ${\psi _i^{\left( j \right)}}$ satisfying (\ref{kktcon}) for point ${{\mathbf{p}^ * }}$, then $f\left( {{\mathbf{p}^ * }} \right)$ must be the  global maximum.

\subsection{The decision function for adaptive link selection}
\begin{proposition}
The global optimal decision function is
\begin{equation}
\begin{array}{l}
p_0^*(i)=1,~~ {\rm{when}}~ - \frac{{\lambda  + 1}}{{\lambda  + \mu  + 1}} \le \frac{{{C_0}\left( i \right)}}{{{C_2}\left( i \right)}}\\
p_1^*(i)=1,~~ {\rm{when}}~ - \frac{{\lambda  + \mu  + 1}}{{\mu  + 1}} \le \frac{{{C_0}\left( i \right)}}{{{C_2}\left( i \right)}} \le  - \frac{{\lambda  + 1}}{{\lambda  + \mu  + 1}}\\
p_2^*(i)=1,~~ {\rm{when}}~ \frac{{{C_0}\left( i \right)}}{{{C_2}\left( i \right)}} \le  - \frac{{\lambda  + \mu  + 1}}{{\mu  + 1}}
\end{array} .
\label{decision}
\end{equation}
\end{proposition}
The proof is based on KKT conditions (\ref{kkt1})-(\ref{kkt5}) and provided in Appendix A. Next, we will use the KKT condition (\ref{kkt6}) to get the thresholds $\lambda ,~\mu $ in the decision function.

The KKT condition (\ref{kkt6}) is equivalent to (\ref{p0p1p2s2}). The probability density function for ${{\gamma _l}\left( i \right)}$ is $f\left( {{\gamma _l}} \right) = \frac{1}{{{\Omega _l}}}{e^{ - \frac{{{\gamma _l}}}{{{\Omega _l}}}}}, {\gamma _l} > 0$,~$l=1,2$. From (\ref{decision}), we have
\begin{small}
\begin{eqnarray}
\label{integral1} \nonumber &E\left\{ {{p_0}(i){C_0}\left( i \right)} \right\} {=} \int\limits_0^\infty  {\left[ {\int\limits_{{L_1}}^\infty  {{{\log }_2}\left( {1 {+} {\gamma _0}} \right)} f\left( {{\gamma _0}} \right)d{\gamma _0}} \right]} f\left( {{\gamma _2}} \right)d{\gamma _2}\\
\label{integral2} \nonumber &E\left\{ {{p_2}(i){C_2}\left( i \right)} \right\} {=} \int\limits_0^\infty  {\left[ {\int\limits_{{L_2}}^\infty  {{{\log }_2}\left( {1 {+} {\gamma _2}} \right)} f\left( {{\gamma _2}} \right)d{\gamma _2}} \right]} f\left( {{\gamma _0}} \right)d{\gamma _0} \\
\label{integral3} \nonumber &E\left\{ {{p_1}(i){C_2}\left( i \right)} \right\} {=} \int\limits_0^\infty  {\left[ {\int\limits_{{L_3}}^{{L_2}} {{{\log }_2}\left( {1 {+} {\gamma _2}} \right)} f\left( {{\gamma _2}} \right)d{\gamma _2}} \right]} f\left( {{\gamma _0}} \right)d{\gamma _0}\\
\label{integral4}  &E\left\{ {{p_1}(i){C_0}\left( i \right)} \right\} {=}  \int\limits_0^\infty  {\left[ {\int\limits_{{L_4}}^{{L_1}} {{{\log }_2}\left( {1 {+} {\gamma _0}} \right)} f\left( {{\gamma _0}} \right)d{\gamma _0}} \right]} f\left( {{\gamma _2}} \right)d{\gamma _2}
\end{eqnarray}
\end{small}
here ${L_1} = {\left( {{\gamma _2} + 1} \right)^{ - \frac{{\lambda  + 1}}{{\lambda  + \mu  + 1}}}} - 1$, ${L_2} = {\left( {{\gamma _0} + 1} \right)^{ - \frac{{\mu  + 1}}{{\lambda  + \mu  + 1}}}} - 1$, ${L_3} = {\left( {{\gamma _0} + 1} \right)^{ - \frac{{\lambda  + \mu  + 1}}{{\lambda  + 1}}}} - 1$, ${L_4} = {\left( {{\gamma _2} + 1} \right)^{ - \frac{{\lambda  + \mu  + 1}}{{\mu  + 1}}}} - 1$. Substituting (\ref{integral4}) into (\ref{p0p1p2s2}), and simplifying the integral equations using (\ref{sim}), we can find a numerical solution for $\lambda$ and $\mu$, which are used in section \ref{numeric} to determine the sum rate.
\begin{equation}
\begin{array}{l}
\int\limits_a^b {\ln \left( {1 + x} \right)\frac{{{e^{ - \frac{x}{\Omega }}}}}{\Omega }} dx = {e^{ - \frac{a}{\Omega }}}\ln \left( {1 + a} \right) - {e^{ - \frac{b}{\Omega }}}\ln \left( {1 + b} \right)\\
 {+} {e^{\frac{1}{\Omega }}}\left\{ {{E_1}\left( {\frac{{a + 1}}{\Omega }} \right) {-} {E_1}\left( {\frac{{b + 1}}{\Omega }} \right)} \right\},~~{E_1}\left( z \right) = \int\limits_z^\infty  {\frac{{{e^{ - t}}}}{t}dt}.
\end{array}
\label{sim}
\end{equation}

\section{Reference system}
The transmission schedule of the reference system is fixed and determined in advance.
The transmissions from U$0$, U$2$ and RS$1$ are done sequentially and
have durations  $t_0$, $t_2$ and $t_1$ respectively. The transmissions span over many channel fades so that
 each link is assumed to achieve the ergodic capacity. The ergodic capacity of link ${\gamma _l}$ is
\begin{equation}\nonumber
{C_e}\left( l \right) = \int\limits_0^\infty  {{{\log }_2}\left( {1 + {\gamma _l}} \right)} f\left( {{\gamma _l}} \right)d{\gamma _l} = \frac{{{e^{\frac{1}{{{\Omega _l}}}}}}}{{\ln 2}}{E_1}\left( {\frac{1}{{{\Omega _l}}}} \right),~l=0,2
\label{cerg}
\end{equation}
Furthermore, RS$1$ is assumed to be equipped with infinitely large buffers.

The broadcast period is ${t_1} = \max \left\{ {\frac{{{C_e}\left( 0 \right){t_0}}}{{{C_e}\left( 2 \right)}},\frac{{{C_e}\left( 2 \right){t_2}}}{{{C_e}\left( 0 \right)}}} \right\}$. The sum-rate of this system is ${R_{rf}} = \frac{{{C_e}\left( 0 \right){t_0} + {C_e}\left( 2 \right){t_2}}}{{{t_0} + {t_1} + {t_2}}}$. Optimizing w.r.t. ${t_0},{t_1},{t_2}$, the maximal sum-rate of the reference system is $R_{rf}^* = \frac{{{C_e}{{\left( 0 \right)}^2}{C_e}\left( 2 \right) + {C_e}{{\left( 2 \right)}^2}{C_e}\left( 0 \right)}}{{{C_e}{{\left( 0 \right)}^2} + {C_e}{{\left( 2 \right)}^2} + {C_e}\left( 0 \right){C_e}\left( 2 \right)}}$.

\section{Numerical results}\label{numeric}
\begin{figure}[t]
\centering
\includegraphics[width=7.2 cm]{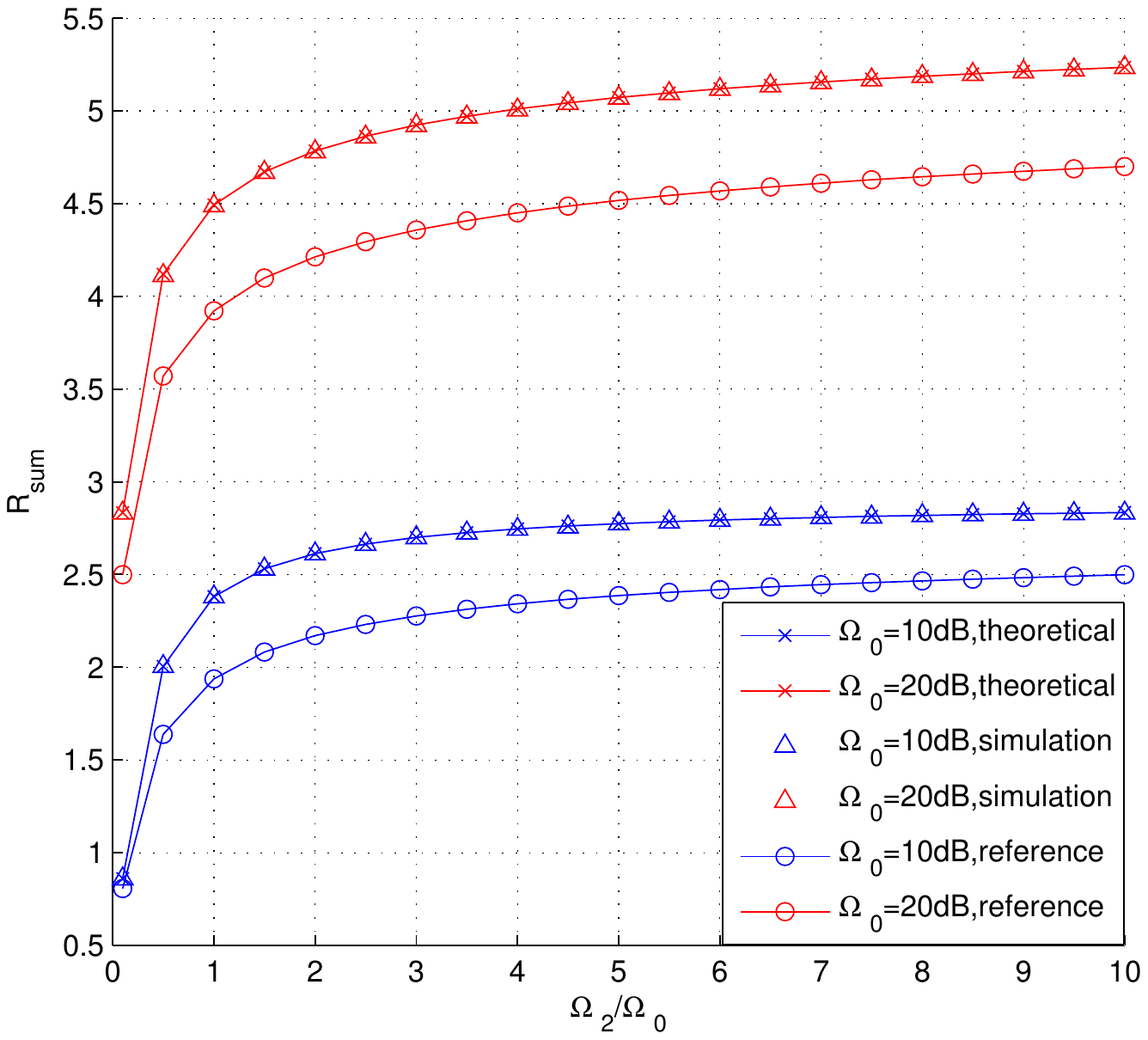}
\caption{Sum-rate under different SNR condition}
\label{result}%
\vspace{-5pt}
\end{figure}

\begin{figure}[t]
\centering
\includegraphics[width=7.3 cm]{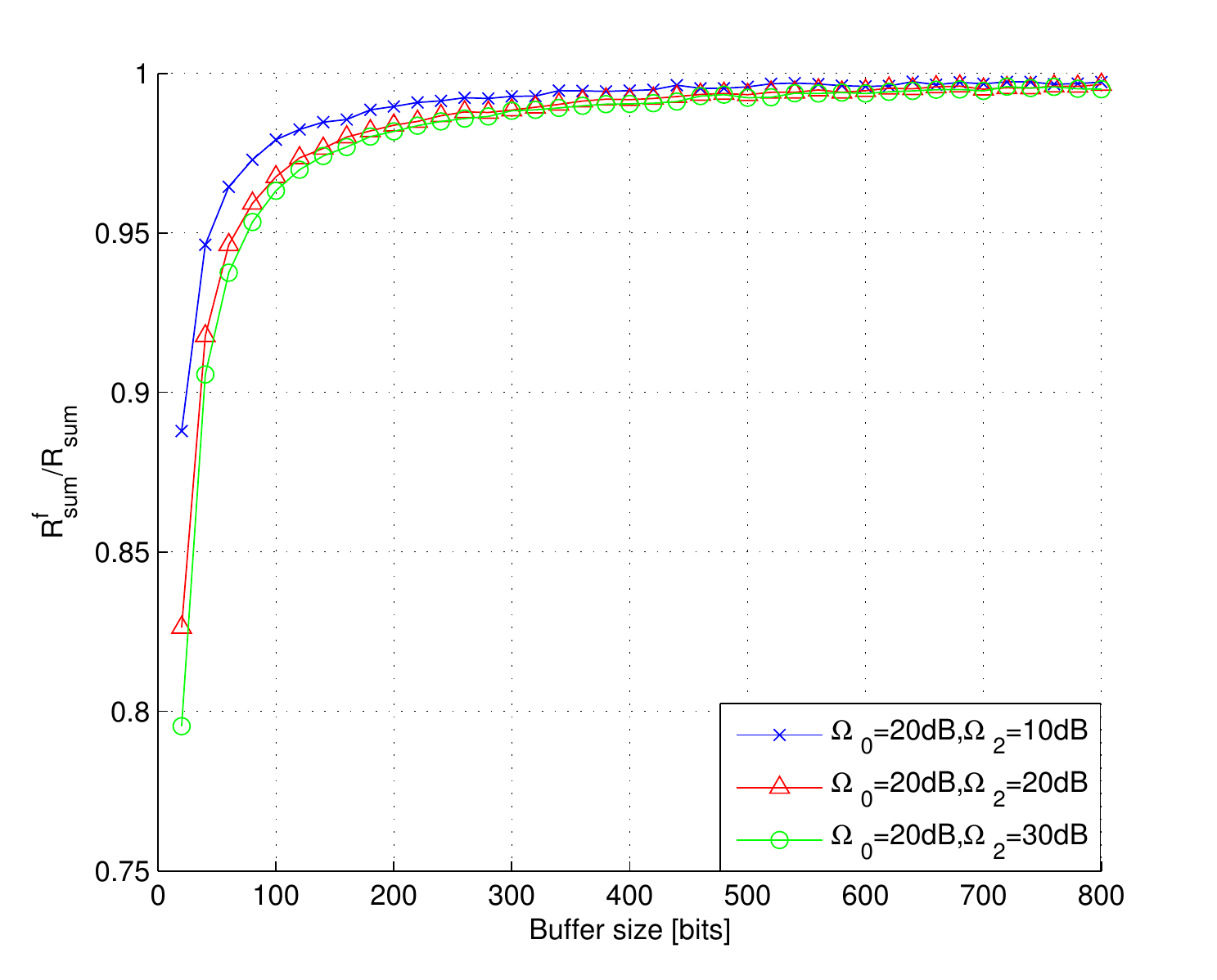}
\caption{Performance of finite buffer size}
\label{finitebf}%
\vspace{-5pt}
\end{figure}

Setting ${\Omega _{\rm{0}}}{\rm{ = 10dB}}$ or ${\Omega _{\rm{0}}}{\rm{ = 20dB}}$ and ${{{\Omega _{\rm{2}}}} \mathord{\left/
 {\vphantom {{{\Omega _{\rm{2}}}} {{\Omega _{\rm{0}}}}}} \right.
 \kern-\nulldelimiterspace} {{\Omega _{\rm{0}}}}}$ from 0.1 to 10, we solve the integral equations (\ref{integral4}) to get the thresholds $\lambda , \mu$ and the theoretical results of the sum-rate shown in Fig.~\ref{result}.
Applying the thresholds $\lambda , \mu$ obtained from the integral equations and the decision function of (\ref{decision}) into the simulation, we get the simulation results of the sum-rate shown in Fig.~\ref{result}. We observe that the theoretical results and the simulation results coincide. Whatever the SNRs ${\gamma _0}$ and ${\gamma _2}$, the sum-rate of the proposed adaptive link selection protocol is larger than the sum-rate of the reference system achieving the ergodic capacity for each single link.

We now show the effect of a limitation in the buffer size. We find the optimal $\lambda , \mu$ from the integral equations and the decision function in (\ref{decision}) but with the following additional constraint: if ${C_l}\left( i \right) + {Q_l}\left( {i - 1} \right) > \max  Q_l$, link $l$ is not chosen.
 The sum-rate for finite buffer size is denoted as $R_{sum}^f$.  Fig.~\ref{finitebf} shows the simulation results as a function of  the buffer size when the  buffer sizes of ${Q_0}$ and ${Q_2}$ are equal.

\section{Conclusion}
We exploit the presence of a buffer at the relay of a TDBC two-way relay network to opportunistically schedule the communication links maximizing the system sum-rate and guaranteeing buffer stability conditions.
We use the KKT approach to derive the decision function for adaptive link selection under a Rayleigh fading assumption.
The numerical results show that the proposed protocol outperforms the reference protocol where scheduling is determined in advance and does not depend on the channel state information. An interesting issue for future work is to investigate two-way relaying with a common buffer of a limited size, as well as the impact of the direct link between the users, along with a compression-and-forward relaying strategy.

\appendices{}
\section{}
The gradient equation (\ref{kkt1}) is equivalent to the following equations for $\forall i$
\begin{subequations}
\vspace{-2pt}
\label{firstordr}
\begin{align}
\label{c0} &- \lambda {C_0}\left( i \right) + {\alpha _i} - \beta _i^{\left( 0 \right)} + \psi _i^{\left( 0 \right)} = 0\\
\label{c2} & - \mu {C_2}\left( i \right) + {\alpha _i} - \beta _i^{\left( 2 \right)} + \psi _i^{\left( 2 \right)} = 0 \\
\label{c0c2} &\left( {\mu  + 1} \right){C_0}\left( i \right) {+} \left( {\lambda  + 1} \right){C_2}\left( i \right) + {\alpha _i} - \beta _i^{\left( 1 \right)} {+} \psi _i^{\left( 1 \right)} {=} 0.
\end{align}
\end{subequations}
In Appendix B, we prove that the solutions ${p_j^*(i)}$ of (\ref{firstordr}) can only be 0 or 1.

We first examine the case $p_0^*(i) = 1$, $p_1^*(i) = 0$, $p_2^*(i) = 0$.
Considering (\ref{kkt3}) and (\ref{kkt4}), we have $\beta _i^{\left( 1 \right)} = \beta _i^{\left( 2 \right)} = 0$ and $\psi _i^{\left( 0 \right)} = 0$ respectively. Then Subtracting (\ref{c0}) from (\ref{c2}) and (\ref{c0c2}), we get
\begin{subequations}
\vspace{-2pt}
\label{eqcon1}
\begin{align}
\label{eqcon1s1} &\lambda {C_0}\left( i \right) - \mu {C_2}\left( i \right) + \psi _i^{\left( 2 \right)} + \beta _i^{\left( 0 \right)} = 0\\
\label{eqcon1s2} &\left( {\lambda  + \mu  + 1} \right){C_0}\left( i \right) + \left( {\lambda  + 1} \right){C_2}\left( i \right) + \psi _i^{\left( 1 \right)} {+} \beta _i^{\left( 0 \right)} {= }0.
\end{align}
\end{subequations}
From (\ref{kkt3}) and (\ref{kkt4}), $\psi _i^{\left( 1 \right)},\psi _i^{\left( 2 \right)},\beta _i^{\left( 0 \right)} \ge 0$ must be satisfied. Hence the necessary conditions for $p_0^*(i) = 1$ are as follow
\begin{subequations}
\vspace{-2pt}
\label{ineqcon1}
\begin{align}
\label{ineqcon1s1} &\lambda {C_0}\left( i \right) - \mu {C_2}\left( i \right) \le 0\\
\label{ineqcon1s2} &\left( {\lambda  + \mu  + 1} \right){C_0}\left( i \right) + \left( {\lambda  + 1} \right){C_2}\left( i \right) \le 0.
\end{align}
\end{subequations}
Obviously if inequalities (\ref{ineqcon1}) are satisfied, we definitely can find appropriate coefficients $\psi _i^{\left( 1 \right)},\psi _i^{\left( 2 \right)},\beta _i^{\left( 0 \right)} \ge 0$ verifying (\ref{eqcon1}).
As the only constraints on ${\alpha _i}$ are contained in  (\ref{firstordr}),  if (\ref{eqcon1}) is satisfied, we can find an appropriate ${\alpha _i}$ to meet (\ref{firstordr}) as well. So (\ref{ineqcon1}) is also the sufficient conditions for $p_0^*(i) = 1$.

Similarly, for $p_2^*(i) = 1$ and $p_1^*(i) = 1$, the equivalent conditions are (\ref{ineqcon2}) and (\ref{ineqcon3}) respectively.
\begin{subequations}
\vspace{-3pt}
\label{ineqcon2}
\begin{align}
\label{ineqcon2s1} &\lambda {C_0}\left( i \right) - \mu {C_2}\left( i \right) \ge 0\\
\label{ineqcon2s2} &\left( {\mu  + 1} \right){C_0}\left( i \right) + \left( {\lambda  + \mu  + 1} \right){C_2}\left( i \right) \le 0.
\end{align}
\end{subequations}
\vspace{-18pt}
\begin{subequations}
\label{ineqcon3}
\begin{align}
\label{ineqcon3s1} &\left( {\lambda  + \mu  + 1} \right){C_0}\left( i \right) + \left( {\lambda  + 1} \right){C_2}\left( i \right) \ge 0\\
\label{ineqcon3s2} &\left( {\mu  + 1} \right){C_0}\left( i \right) + \left( {\lambda  + \mu  + 1} \right){C_2}\left( i \right) \ge 0.
\end{align}
\end{subequations}

Next, we simplify conditions (\ref{ineqcon1}-\ref{ineqcon2}) by removing the redundant equations.
From (\ref{ineqcon1}), (\ref{ineqcon2}) and (\ref{ineqcon3}), we can infer ${\rm{ - 1}} \le \lambda , \mu  \le {\rm{0}}$ and $\lambda  + \mu  + 1 \le {\rm{0}}$. The proof is deferred to Appendix C. By cancelation ${C_0}\left( i \right)$ and ${C_2}\left( i \right)$ in (\ref{ineqcon3}) we obtain
\begin{equation}
\vspace{-4pt}
{\lambda ^2} + {\mu ^2} + \lambda \mu  + \lambda  + \mu  \le 0.
\label{mulam}
\end{equation}

$\lambda , \mu$ are constant parameters (do not depend on time index $i$), so (\ref{mulam}) should also be valid in conditions (\ref{ineqcon1}) and (\ref{ineqcon2}). Hence, when
(\ref{ineqcon1s2}) is verified, so is (\ref{ineqcon1s1}). Similarly, when (\ref{ineqcon2s2}) is verified, so is (\ref{ineqcon2s1}).

Hence, we have proven that if (\ref{decision}) is valid, we can find appropriate coefficients $\psi _i^{\left( j \right)},\beta _i^{\left( j \right)} \ge 0$ and ${\alpha _i}$ to meet the KKT condition (\ref{kkt1})-(\ref{kkt5}).
\section{}
If there exists one non-binary number among ${p_0^*(i)}$, ${p_1^*(i)}$ and ${p_2^*(i)}$, there are at least two non-binary numbers among ${p_0^*(i)}$, ${p_1^*(i)}$ and ${p_2^*(i)}$, otherwise ${{p_0^*(i)}  + {p_1^*(i)}  + {p_2^*(i)}} = 1$ can not be satisfied. We assume ${p_0^*(i)}$ and ${p_2^*(i)}$ are non-binary for instance, the proof for other cases is similar. Then ${p_1^*(i)}$ can not be 1. From (\ref{kkt3}) and (\ref{kkt4}),  we obtain $\beta _i^{\left( 0 \right)} = \beta _i^{\left( 1 \right)} = \beta _i^{\left( 2 \right)} = 0$ and $\psi _i^{\left( 0 \right)} = \psi _i^{\left( 2 \right)} = 0$ respectively. Substituting the above results into (\ref{c0}) and (\ref{c2}) which are valid for $\forall i$, we get $\lambda  = \mu  = {\alpha _i} = 0$. Then considering (\ref{c0c2}), we obtain ${C_0}\left( i \right) + {C_2}\left( i \right) + \psi _i^{\left( 1 \right)} = 0$. However, ${C_0}\left( i \right),{C_2}\left( i \right) > 0$ and $\psi _i^{\left( 1 \right)} \ge 0$ should be satisfied, which is a contradiction. So ${p_0^*(i)}$, ${p_1^*(i)}$ and ${p_2^*(i)}$ are on the border of $[0,1]$.
\section{}
Note that ${C_0}\left( i \right),{C_2}\left( i \right) > 0$ is always satisfied. $\lambda$ and $\mu $ must have the same sign, otherwise (\ref{ineqcon1s1}) or (\ref{ineqcon2s1}) will be valid constantly  which means U$0$ or U$2$ will transmit in every time slot. $\lambda , \mu  \le {\rm{0}}$ must be satisfied, otherwise (\ref{ineqcon1s2}) and (\ref{ineqcon2s2}) will never happen which means RS$1$ will broadcast in every time slot. And ${\rm{ - 1}} \le \lambda , \mu $ also should be satisfied, otherwise (\ref{ineqcon3}) will never happen which means there is no broadcast at all. So we get ${\rm{ - 1}} \le \lambda , \mu  \le {\rm{0}}$. Similarly, $\lambda  + \mu  + 1 \le {\rm{0}}$ is valid, otherwise RS$1$ will broadcast in every time slot.

\bibliographystyle{IEEEtran}
\bibliography{hua}
\end{document}